# The problem of the relationship qualitative data, quantitative data in general statistics

El problema de la relación datos cualitativos, datos cuantitativos en la estadística general
Jhon Jairo Mosquera Rodas https://orcid.org/
0000-0002-3455-8470
The author, Jhon Jairo Mosquera Rodas. Worked at the Universidad Cooperativa de Colombia, Pereira, Rda. 660007 Colombia. He now works at the Department of Psychology of the Universidad Cooperativa de Colombia, attached as a researcher in Mathematics and Physics (e-mail: jhon.mosquera@campusucc.edu.co).

*Abstract*— The disjunction between nature and science is studied, together with the need to modify the conception of natural science vs artificial science, related to the perspective of objectivity and subjectivity, to end with the explanation of the process of polarization of methodologies and the relationship between mixed data, as a possibility of unification of qualitative and quantitative data, through relationships and correlations.

*Index terms*— Qualitative, quantitative, knowledge, complexity.

Resumen— Se estudia la disyuntiva entre naturaleza y ciencia, junto con la necesidad de modificar la concepción de ciencia natural vs. ciencia artificial, relacionada con la perspectiva de objetividad y subjetividad, para terminar con la explicación del proceso de polarización de las metodologías y la relación entre los datos mixtos, como posibilidad de unificación de los datos cualitativos y cuantitativos, a través de relaciones y correlaciones.

Palabras claves— Cualitativo, cuantitativo, conocimiento, complejidad.

## I. INTRODUCTION

The research raises the dilemma between nature and science, focusing on the need to modify the conception of natural science vs. social science, proposing a different analysis of the concepts of objectivity and subjectivity in the face of the exercise of polarisation of methodologies throughout history and the mixed relationship that emerges as a possibility of linking the researcher-culture relationship, differentiating it from the classic subjectivity vs. objectivity relationship.

On the other hand, it analyses the three definitions that imply duality, together with the criterion of unity in scientific duality, through the analysis of the data in quantitative research and its relationship with the data in qualitative research, the treatment of the same, using the possible correlations together with the first approach to the technique of effectual linking, which allows linking in a whole both types of data, to assume a type of methodology that admits linking such relationships and correlations, which are identified as natural both in research in natural sciences and in social sciences.

## II. MATERIALS AND METHOD

Method: The research is of an exploratory type according to its scope, as it carries out an interdisciplinary analysis of little known topics, showing the relationship between qualitative and quantitative data, seen as a whole that can communicate through relational and correlational analysis, taking into account the different tensions that arise between the data when manipulated by the researcher, thus applying the first step of the technique of effective linking, to generate a relevant and precise analytical process in terms of the solution to the research problem. The approach is qualitative, as the research is characterised by observations and descriptions of the phenomenon of relationships and correlations between different types of data.

## III. THE DISJUNCTION BETWEEN NATURE AND SCIENCE.

Man could have taken one of two paths at the beginning of the modern scientific paradigm, the path of reason without context or the path of natural reason: for obvious reasons the founders of western science opted for the path of reason without natural context, generating multiple schools that supported this scientific position, leaving behind the natural conception that implies the constant relationship with the context, i.e., we stopped experimenting in the natural and started experimenting only in the laboratory, this factor determines a type of artificial scientific paradigm, to which Sanguineti alludes when analysing the inference of Thomas Aquinas.

"The world is one not because it has one nature - it is a unity of order, not a substance - but because the beings that compose it act naturally in relation to one another. The nature of natural things," says Thomas Aquinas, "is the operative principle of each entity in what pertains to it in relation to the order of the universe. Its intrinsic and spontaneous character distinguishes it from the artificial order established by man". (Sanguineti, 2009, p. 2008). [13]

The need to modify the conception of natural science vs artificial science

It is not necessary to make an analysis of the development of artificial science, which Thomas Aquinas names as the artificial order, which by its own advances has made great developments at the scientific and technological level, using reason in the hypothetical or artificial context, as the case may be. But the paradigm must be renewed and this is something that currently points to complexity as a fundamental element for the survival of the human race.

"This is where the idea of complexity comes in, and the need for a reform of thought, which allows us to access a general theory that integrates the various academic disciplines and opens us to a more humane and supportive attitude". (Morin 2015, p. 208) [9]

"We are living today what one author has graphically called "specialised deafness", which is nothing more than the loss of "generalised hearing", that is, the inability to communicate, a situation more common in academic life, which needs mutual encounter and criticism in order to be able to live" (Morin 2015 p. 208). [9]

This means that, according to Morin, a transformation of the current scientific paradigm is necessary in order to move towards forms of rationality that are closer to the human, devoid of the artificial reason without context that modern science inaugurates, and which today is in crisis, as it does not respond adequately to the challenges posed by contemporaneity. It is necessary to move from formal scientific discourse to formal scientific action, focused on the development of the ideal conditions for the generation of knowledge that is necessary, effective, but above all relevant to the needs of the generations that precede us. Moving away from the conception of artificial science vs. natural science is the step to follow in order to transform the global landscape through science.

Objectivity and Subjectivity in Science

The structural division of science into a qualitative and a quantitative block generates complexities when carrying out common analyses through the statistical method. This type of statistical and methodological ambiguity is solved through the proposal of the mixed methodology provided by different aspects, as Pole points out below:

"The historical debate around qualitative and quantitative research paradigms has at times been passionate. Arguments for and against the methodologies have often focused on philosophical differences with respect to issues such as generalisability, epistemology or the authentic representation of the phenomena under investigation. Recently, however, a considerable focus of discussion has centred on how mixed-method research can be conducted effectively. Broadly speaking, mixed methodologies can be conceptualised as the use or combination of research methodologies from both quantitative and qualitative traditions." (Pole, 2009 p. 37) [11]

This kind of union between the extremes generates the possibility of harmonising qualitative and quantitative processes, depending on the resolution of the problem question that the researcher has previously defined.

"It is also recognised that all data collection, quantitative or qualitative, operates within a cultural context and is affected by the tendencies and beliefs of the collectors. As Anthony J. Onwuegbuzie (personal communications, January 30, 2005) puts it: "(Cited by Pole, 2009 p. 39) [11]

The relationship that Pole points out in the previous paragraph with regard to the triad quantitative data-qualitative data-culture is relevant, which goes beyond the subjective incidence of the data collectors, becoming a differentiating element of the quality of the data together with what is understood as objective or subjective on the part of the researcher. It is worth noting that the tendency to see in subjective data a certain level of falsity is inherited from the objective conception of science coming from the quantitative block of the natural sciences and from the philosophical materialist schools.

"As qualitative methodology rapidly gained popularity among some researchers, they began to engage in so-called paradigm wars (Gage, 1989), [5] with each end of the quantitative/qualitative argument criticising the methods, procedures and validity of the other's results. These paradigm wars served to polarise both sides of the debate."(Cited by Pole, 2009 p. 39). [11]

This aspect pointed out by Pole is the fundamental element of research, since the division of science into two blocks has generated the fragmentation of knowledge, preventing us from seeing the whole in its parts and the parts in the whole, in a constant and non-fragmented relationship.

"Mixed-method research can provide more robust inferences because data are viewed from multiple perspectives. One method may provide greater depth, the other greater breath, and together they confirm or complement each other. For example, quantitative data can be used to measure the success of an intervention and qualitative data to explain the process of the intervention. Mixed methodologies are useful when they offer better opportunities to answer the research questions of interest and when they help the researcher to assess how correct their ideas are (Tashakkori and Teddlie, 2003: 14)". (Quoted by Pole, 2009 p. 39) [11]

Here the issue of data is seen by Tashakkori and Teddlie from a distinctly mixed perspective, and it is important to note that this focus on both methodologies and the use of data is clearly a significant advance for 21st century science.

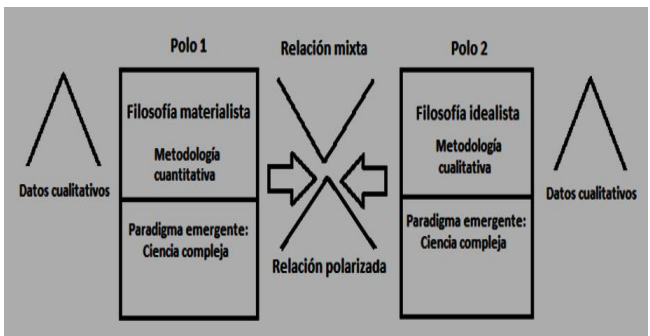

Figure 1. Polarisation of methodologies and the mixed relationship.

What is remarkable in figure 1 is the mixed relationship that emerges from the appearance on the science scene of the complex paradigm; this aspect generates processes of accommodation between quantitative and qualitative models of science, giving rise to mixed relationships, the fruit of the contribution given by complexity to the system.

"There are also circumstances, although I consider them rare, in which two paradigms can coexist peacefully in the late period. (Kuhn, 2004 p. 16) [8]

Kuhn makes here a key discovery for science, he states that there is a context in which different paradigms of science coexist and that this process generates frictions between these, the really important thing is in the possibility that two paradigms coexist, aspect that happens at present, since the classic paradigm of the normal science and the emergent paradigm of the complexity, generate a process of key transformation for the current scientific development. But Kuhn forgets to describe the same process in depth and the relations that are carried out in this context, being one of these, the mixed relation that raises the coexistence between the quantitative and the qualitative in science, deriving from the previous thing a methodology that obeys to the same principles.

IV CLASSIFICATION OF INFORMATION

Data in quantitative research

The information in quantitative research organises the data in a rigorous way, generating the impression of greater objectivity, by using the numerical relation, together with the interpretation. Muñoz Campos refers to it as follows.

"According to (Muñoz Campos, 2000:15, citado por Aguilar, 2014. p. 86), [1] the main characteristics of quantitative research are: application of the hypothetical deductive method, representation of representative samples, objective measurement of variables, use of quantitative data collection techniques with instruments such as questionnaires, scales, tests, statistical application in data analysis and the testing of hypotheses and theories"

Whereas in quantitative research exactly the opposite is true.

Data in Qualitative Research

It is important to highlight that qualitative research has elements typical of the inductive method, and that due to the nature of this method, it will tend to be limited to the field of in-depth study or description of the phenomena that the researcher wants to deal with.

"(...) Six elements are fundamental to pose a qualitative problem: research objectives, research questions, research justification, research feasibility, evaluation of the deficiencies in the knowledge of the problem and initial definition of the environment or context. (Hernández, Fernández and Baptista, 2014. p. 356). [7]

These being the essential elements, with the hypothesis, it is important to identify the characteristics of the data, Ryan and Bernard, make this contribution, as follows.

"When we talk about qualitative data we generally talk about texts: newspapers, films, comedies, e-mail messages, stories, life stories; and also narratives, such as, for example, stories about divorce, about being ill (...) (Ryan and Bernard, 2003, cited by Fernández, 2006). [4]

In the following, Ryan and Bernard list the different phases in the process of data analysis from the qualitative approach.

"1. Obtaining information: through the systematic recording of field notes, obtaining documents of various kinds, and conducting interviews, observations or focus groups. (...).

2. Capturing, transcribing and ordering the information: the information is captured through various means. Specifically, in the case of interviews and focus groups, through electronic recording (recording on cassettes or in digital format). (...).

3. Coding the information: Coding is the process by which the information obtained is grouped into categories that concentrate the ideas, concepts or similar themes discovered by the researcher, or the steps or phases within a process (Rubin and Rubin, 1995, citado por Fernández, 2006.p. 4). (...). [4]

4. Integrate the information: relate the categories obtained in the previous step to each other and to the theoretical foundations of the research. (...). (Fernández, 2006.p. 4). [4]

Classification of the information

The way of classifying information is different from the quantitative research approach. Fernandez refers to some methods to carry out such a process.

"There are several methods for collecting and analysing words or phrases. Data collection techniques include free lists, stack sorting, paired comparisons, triad testing and structure substitution tasks. Techniques for analysing these types of data include componential analysis, taxonomies and mind maps". (Fernandez, 2006, p. 3). [4]

The Data: Essential Elements of Statistical Analysis

Here it is necessary to point out the difference between data and their respective relationship with statistics, and by making this distinction and the corresponding typology, the question of the correspondence between qualitative data and quantitative data can be better understood. Data is then defined as:

"Traditionally data has been defined as a symbol that has not yet been interpreted according to Spek and Spijkervet (1997, p. 20), a simple observation of the state of the world according to Davenport (1997, p. 45), or as a raw, simple and discrete fact as stated by Bhatt (2001, p. 72), Beveren (2002, p. 19), Davenport and Prusak (1998, p. 46) and Herder et al. (2003, p. 110). (Cited by Arias and Aristizabal, 2011, p. 98). [3]

The aforementioned authors present in the very nature of data, the distinction between qualitative and quantitative data; the former are based on the observation of the phenomenon and the latter are like a symbol not elaborated by the researcher through interpretation, which is the beginning of the research process in natural and social science. Gil presents an approach to the concept of data, taking into account the mediation of the researcher and the use of interpretation.

"Most authors assume that the researcher plays an active role with respect to the data: the data is the result of the elaboration process, i.e. it has to be constructed" (Gil, 1994, p.25). (Gil, 1994, p.25) [6]

The data is then presented as a product of the collection of information, manipulated by the researcher and organised according to its nature, in a qualitative or quantitative structure, as the case may be. Murdick goes on to define data, highlighting its formal character.

"They are a basic set of facts concerning a person, thing or transaction. They include such things as size, quantity, description, volume, rate, name or place. (Murdick, p. 157). [10]

The above characteristics become increasingly specific according to the approach the researcher employs.

"Quantitative data
Data that can be measured or quantified (counted).
Qualitative data
Data that cannot be expressed numerically. They represent a quality or attribute that classifies each subject in one of several categories". (Araujo, 2011. Para.5,13) [2]

This distinction provided by Araujo, although it has been useful for science, has also preserved in the system the dualistic position as a central element of the statistical process, to the point of the historical existence of the categories hard sciences and soft sciences that evidences the dual movement in the research processes, under the imprint of a more reliable type of scientific approach than its opponent. The following is a description of the types of data used in both approaches.

V. CORRELATIONS AND TENSIONS BETWEEN QUALITATIVE AND QUANTITATIVE DATA

In the following, an attempt will be made to link the types of data from both methodological blocks, uncovering emerging relationships, through specific correspondences.

TABLE I.
CORRELATIONS DATES

| Características de los datos cualitativos | Correspondencia unívoca | Correspondencia no unívoca | Correspondencia inversa |
|---|---|---|---|
| a. Expansión | ---------- | ---------- | 1 |
| b. desproporción | ---------- | ---------- | 1 |
| c. Imprecisión | ---------- | ---------- | 1 |

The first classification corresponds to the inverse relationship the data are still presented in opposites:

Pair 1: a. expansion (D. qualitative) ≥ ≤ a1. Acotation (qualitative D.) = inverse correspondence.

Pair 2: b. disproportion (Qualitative D.) ≥ ≤ b1. Measurement (qualitative D.) = Inverse correspondence.

Pair 3: c. Imprecision (qualitative D.) ≥ ≤ c1. Precision (qualitative D.) = Inverse correspondence.

Correspondences obey processes of permanent relationship between qualitative and quantitative data. This in itself is a relationship that has a mechanism of duality and tensions between opposites, which we will call a field of quantitative-qualitative or qualitative-quantitative tensions, depending on the initial focus of the research.

Table 4 presents a field of tensions between quantitative and qualitative data that can be used to generate processes of organising the variables in a possible scenario for the solution of a problem. Only taking into account the inverse correspondence relationship found.

TABLE II.
CORRELATIONS BETWEEN DATA

| *D.c* |  |  |  |  |
|---|---|---|---|---|
| C.I | 1(a1) | 1(b1) | 1(c1) |  |
| a | 1(a(**a1**)) | 1(a(**b1**)) | 1(a(**c1**)) |  |
| b | 1(b(**a1**)) | 1(b(**b1**)) | 1(c(**c1**)) |  |
| c | 1( c( **a1**)) | 1(c(**b1**)) | 1(c(**c1**)) |  |
| ~ | a1. | b1 | c1 | *D:c* |

| *D.c* |  |  |  |  |
|---|---|---|---|---|
| C.I | 1(a1) | 1(b1) | 1(c1) |  |
| a | 1(a(**a1**)) | 1(a(**b1**)) | 1(a(**c1**)) |  |
| b | 1(b(**a1**)) | 1(b(**b1**)) | 1(c(**c1**)) |  |
| c | 1( c( **a1**)) | 1(c(**b1**)) | 1(c(**c1**)) |  |
| ~ | a1. | b1 | c1 | *D:c* |

Source: own elaboration.

| Correspondencia biunívoca | Correspondencia no biunívoca | Características de los datos cuantitativos |
|---|---|---|
| ---------- | ---------- | **a1.** Acotación |
| ---------- | ---------- | **b1.** Medición |
| ---------- | ---------- | **c1.** Precisión |

Source: own elaboration.

This relationship is given by an extreme correspondence between the qualitative data D.c, and the quantitative data D:c, establishing the field of tensions between the data, in a possible scenario.

The row C.I inverse correspondence shows the relations between the elements a, b, c and the elements a1, b1 and c1, in opposite relation between each of the qualitative and quantitative elements. I feel the opposite correlations and expressed as follows: 1(a1) ~ 1(b1) ~1(c1).

The following relationships are the most basic and are expressed as follows:

Row $a^2$ ~ gives the relationships between a~a1, a~b1, a~c1.
Row b ~ poses the relationships between b~a1, b ~b1, b ~c1.
Row c ~ poses the relationships between c~a1, c ~b1, c ~c1.

There is now an inverse relationship that is established from the quantitative data axis, i.e. the column a1~a1, a1 ~b1, a1 ~c1:

Column $a1^4$ ~ posits the relationships between a1~a1, a1 ~b1, a1 ~c1.
Column b1 ~ poses the relationships between b1~a, b1 ~ b1 ~ b1 ~c1.
Column c1 ~ shows the relationships between c1~a, c1 ~b, c1 ~c1.

Here are the relations that exist between the opposites qualitative data, quantitative data. It is similar to the infinite number that appears between 2 rational, irrational or real numbers in mathematics. But expressed in tangential relationships and made possible by statistics, at a level more applicable to quantitative or qualitative analysis processes.

**1/4**……………………………………………**2/4**

TABLE III
CORRELATIONS BETWEEN DATA IN THE STRESS FIELD

| D.c | | | | |
|---|---|---|---|---|
| C.I | 1(a1) | 1(b1) | 1(c1) | |
| a | 1(a(a1)) | 1(a(b1)) | 1(a(c1)) | |
| b | 1(b(a1)) | 1(b(b1)) | 1(c(c1)) | |
| c | 1( c(a1)) | 1(c(b1)) | 1(c(c1)) | |
| ~ | a1. | b1 | c1 | D:c |

| D.c | | | | |
|---|---|---|---|---|
| C.I | 1(a1) | 1(b1) | 1(c1) | |
| a | 1(a(a1)) | 1(a(b1)) | 1(a(c1)) | |
| b | 1(b(a1)) | 1(b(b1)) | 1(c(c1)) | |
| c | 1(c(a1)) | 1(c(b1)) | 1(c(c1)) | |
| ~ | a1. | b1 | c1 | D:c |

Source: own elaboration.

Multiple correlations then emerge between the relationships a~ a1 as well as between the other opposites, which appear in both tables, and these in turn between the pair of main opposites D.c qualitative data, D:c quantitative data. The dividing line tends to become clearer, when the correlations that exist and that ultimately generate the connections between the qualitative data and the quantitative data, expressed in two ways, are progressively found:

Where 1 would be the main relationship: expressed as D.c D:c

Where 1 would be the main ratio: expressed as **D.c $\cong^1$ D:c**

Shows the 3 types of initial correlations expressed by the ~ sign

a. ~1(a(a1))
b. ~1(a(a1))
c. ~1(a(a1))

The difference between the correlations lies in the type of dual correspondence that is generated in item a. (marked in purple) which in mathematics would correspond to the first degree equation and a triadic correspondence generated in item

---
[1] This symbol $\cong$ represents the correlations that exist between two or more general relationships.

b. (marked with green colour), between each of the elements of the correlation. Which in mathematics would correspond to the equation of second degree.

The third combination is quadric, which in mathematics would correspond to the third degree equation. (Marked with brown colour) which in mathematics would correspond to the third degree equation.

These relationships allow for multiple combinations in all directions generating multiple correlations that generate the tension field from the interaction of qualitative and quantitative data, in specific areas. Figure 4 shows the diversity of relationships and correlations in a broader picture.

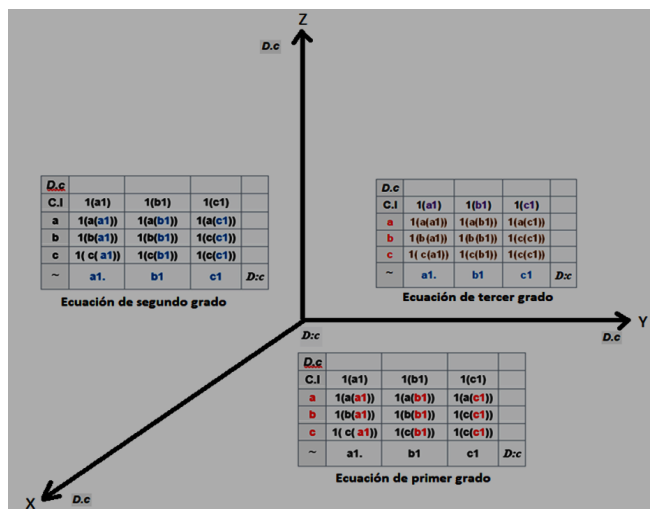

Figure 2. Overview of relationships and correlations between qualitative and quantitative data.

The three minimal ways of relating the correlations, and their respective correspondence with the equations of first, second, third degree and equations of n... degrees allow to see mathematically (by means of the correspondence with the equations of degree) the matter of the relations and the interaction between each one of the types of correlations that can be extended indefinitely, when they are combined with each other.

Now it is necessary to represent the minimal relations in the Cartesian plane taking into account the above distribution.

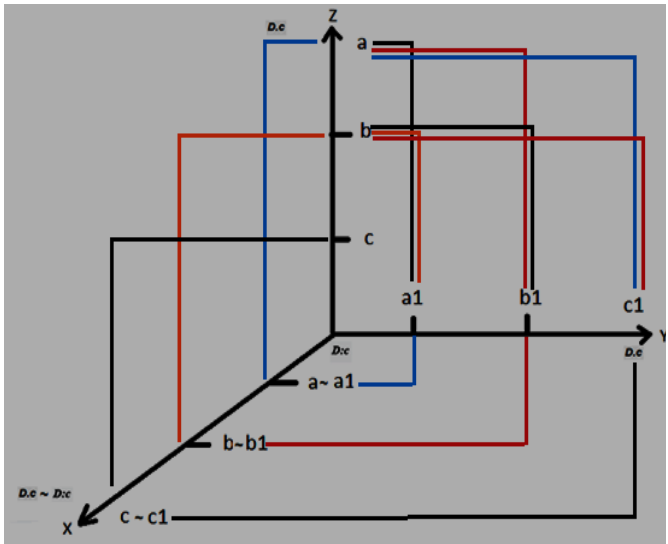

Figure 3. Representation of the relationships that emerge and first perspective of the field of quantitative or qualitative tensions.

The figure plots the first of the multiple fields that emerge from the correlations established between each of the data presented in a possible scenario. These combinations generate permanent or transient connections between D.c and D: c. If analysed closely, the graph specifies specific quadrants or zones that correspond to the elements that constitute the field of tensions between the correlations. These in turn are composed of a multiplicity of relations that generate fields, and these in turn generate other zones that also produce new fields. Thus the centripetal movement internally generates a dynamic that always fluctuates towards the quantitative data, whereas if the centrifugal movement, which belongs to the qualitative data, is followed, the same phenomenon occurs, but in the opposite relation. Thus the dual polarised relations are transformed into dual relations of centripetal and centrifugal movement, but without becoming polarised relations that prevent permanent communication between the parts of the system.

VI. QUALITATIVE DATA TYPES RELATED TO GEOMETRIC SHAPES AND ABSTRACTIONS.

Now, the very nature of quantitative data and this adaptation initiated by the British Association for the Advancement of Science, starting from Stevens' conception, shows not only the adaptation itself, but also the reverse relationship, which starts from objective abstract qualitative data, words that can be verified in relation to exact numbers, based on quantitative elements that can be counted.

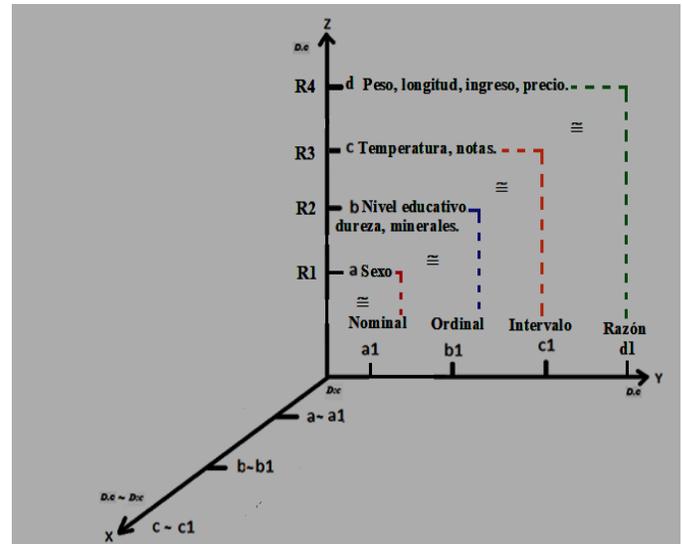

Figure 4. Cartesian plane representation of correlations between qualitative-quantitative data.

On the Cartesian plane one can see more clearly the relationships that exist expressed in quantitative data~ quantitative data; being of a general character, they imply correlations that refer to the various correlational possibilities that exist between opposites.

Here it is important to note that it is becoming more and more evident that the duality between different apparently **opposite** elements is expressed in degrees of approaching and moving away from the point of equilibrium, i.e. science today oscillates in the polarity between opposites, but this is not necessarily the only possibility of interaction between them. The following figure expresses this process.

**Conclusions**

Polarisation of methodologies in quantitative research processes in the different disciplines prevents statistics as an interdisciplinary discipline from developing its full potential in science, and this is an aspect that must be re-evaluated from a logical-mathematical and epistemological perspective.

It is necessary that the field of quantitative and qualitative tensions and their impact throughout history be re-evaluated in relation to the consequences for the scientific method, and the felt need for the integration of the qualitative and the quantitative in a relational order.

Today more than ever, the scientific duality applied to data must be overcome, for which research proposes a defined set of mathematical relationships based on quantitative and qualitative correlations in the Cartesian plane, with their application in descriptive, correlational and explanatory statistics.